\documentclass[%
 aip,
 jmp,
 amsmath,amssymb,
preprint,%
]{revtex4-1}
\usepackage{color}
\usepackage{graphicx}% Include figure files
\usepackage{dcolumn}% Align table columns on decimal point
\usepackage{bm}% bold math
\begin{document}

\title{Defect Analysis of  MCA Wires}% Force line breaks with \\
%\thanks{Footnote to title of article.}

\author{Anita Kumari}
% \altaffiliation[Also at ]{Nano-Computing Research Group, Electrical Engineering \\ University of South Florida.}%Lines break automatically or can be forced with \\
\author{Javier F. Pulecio}%
% \email{Second.Author@institution.edu.}
\author{Sanjukta Bhanja}
\affiliation{ Nano-Computing Research Group, Electrical Engineering \\
 University of South Florida.%\\This line break forced with \textbackslash\textbackslas
}%

\begin{abstract}
As devices continue to scale, imperfections in the fabrication process will have a more substantial impact on the reliability of a system. In Magnetic Cellular Automata (MCA) data is transferred through the coupling of neighboring cells via magnetic force fields. Due to the size of the switching cells, usually of the order of nanometers or smaller, MCA can be sensitive to inherent fabrication defects such as irregular spacing and non-uniform cell structures. Here we investigate conventional electron beam lithography fabrication defects and present a simulation based study on their effects on information propagation in a wire. The study varies the location of the different types of defects throughout the MCA wire under the influence of a spatial moving clocking field. We demonstrate that with the proposed spatially moving clock the most probable fabrication defects of MCA do not affect the information propagation and the location of the defect does not play a significant role in computation. Thus it is concluded that MCA wires demonstrate significant defect robustness towards realistic electron beam lithography shortcomings.

\end{abstract}

%Use showkeys class option if keyword
                              %display desired
\maketitle

\section{\label{sec:Introduction}Introduction}
It is predicted that CMOS devices will stop scaling sometime around the year 2020 and newer technologies will either co-exist with CMOS or replace it~\cite{ITRS}. Although the future of CMOS is unknown, it is clear that many of these novel devices will exist in an embedded sensor computing scenario. Field-coupled Cellular Automata based computing deals with an alternative paradigm where information propagates due to the mutual interaction between neighboring elements as opposed to flow of electrons between two spatial points. This creates a promise for very dense, high speed, and low power computing ~\cite{power}.
This work explores and exploits single-domain magnetic interaction  of magnetic cellular automata architecture, which is already demonstrated by pioneering efforts of Cowburn et. al. and Imre et. al.~\cite{majority}. Attractive features of MCA are the ease of fabrication and room temperature operation. Moreover, there are a multitude of magnetic sensors, already mature MRAM fabrication and sensing techniques suggest better fusion of integrating magnetic logic into an embedded application scenario.

An essential area of study for Magnetic Cellular Automata is the correct directional flow of information from input to output. In order to accomplish information flow, additional external energy is required, namely a clock. This defect study uses a spatially moving clock, as it has demonstrated the correct flow of information in an ideal anti-parallel MCA wire, as opposed to the other clocking schemes that do not propagate information correctly in long wires. The use of a functioning clock allows for proper defect analysis of an anti-parallel coupled MCA wire. We classify defects based on common electron beam lithography errors we have experienced throughout our fabrication techniques. It is critical to study how these defects affect the reliability of the MCA system and is the primary goal of this work.

Based on our fabrication studies (explained in Section~\ref{sec:fab}), we observe five primary types of defects, (1) Irregular spacing (2) Missing material (3) Missing cell (4) Bulging nanomagnet and (5) Merged neighboring cells. This is due to the inherent noisy environment for the electron-beam, impurities, and charge collection. The missing cell and merged cell defect are less frequent than the irregular spacing, missing material and bulge defects. We analyzed the defects using a micro-magnet simulator (Object Oriented Micro-magnetic Framework (OOMMF)) that solves the Landau-Lifshitz equations accounting for various energies (Zeeman energy, Magneto-static energy, Exchange energy, Anisotropy energy, Demagnetization energy etc). This study is restricted to a single defective cell in an array. To characterize the defects, we considered wire lengths of 8 and 16 nanomagnets. Moreover, we considered the role of the location of defect in an array. Salient observations resulting from this study are: 
\begin{itemize}
\item	both length of an array and location of the defect in an array does not affect the output magnetization. 
\item there is almost no magnetization loss when the errors are masked. 
\item MCA is robust to small spacing irregularities, bulge and missing material (less than 5\%) defects for different length MCA arrays, yielding correct output for all defect locations. 
\item only for missing cell, significant amount of missing material and merge cell defect (in general uncommon), the defective magnet retains the previous value causing error. 
\end{itemize}
\vspace{-0.2in}
\section{\label{sec:review}Review of the work}
One of the pioneering efforts in Field-coupled Cellular Automata computing evolved using
quantum tunneling interactions of electrons in neighboring cell~\cite{lent}, promising phenomenal packing density, and the low power-delay product. In a later development, the molecular form of QCA~\cite{molecular-qca} was proposed which works at room temperature alleviating the main drawback of cryogenic operating temperature requirement of E-QCA. The only problem associated with molecular QCA is the difficulty associated with the self-assembly structure. The strong candidate in the field coupled computing is the magnetism, which solves the problem of cryogenic temperature and self assembly structure. In this framework we tried to explore the field coupled computing with nanomagnets for signal processing.
\subsection{\label{sec:review1}Review on MCA}
\begin{figure}
\includegraphics[scale=0.35]{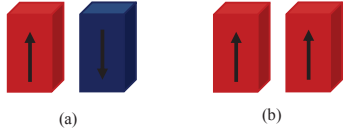}% Here is how to import EPS art
\caption{\label{fig1} (a) Definition of logic '1' and '0' for nanomagnets  (b) Metastable
states for coupled pairs.}
\end{figure}

Magnetic Cellular Automata (MCA) is a variation of the field coupled QCA architecture that was first proposed by Lent et.al.~\cite{lent}. 
 R. P. Cowburn and M. E. Welland~\cite{cowburn2000} proposed a new processing method based on magnetism, in which networks of interacting sub-micrometer magnetic dots are used to perform logic operations and propagate information at room temperature. Later Csaba {\it et.al.},~\cite{Csaba003,Csaba,power} proposed the logic states signaled by the magnetization direction of the single-domain nanomagnet. The nanomagnet coupled to their nearest neighbors through magneto-static interactions, and attempted to align in an anti-parallel ground state as shown in FIG.~\ref{fig1}(a).  FIG.~\ref{fig1}(b) depicts the higher-energy metastable state, which is not desired for correct operation. MCA offers a very low power dissipation~\cite{power} and high integration density of functional devices. 

\begin{figure}
\includegraphics[scale=0.35]{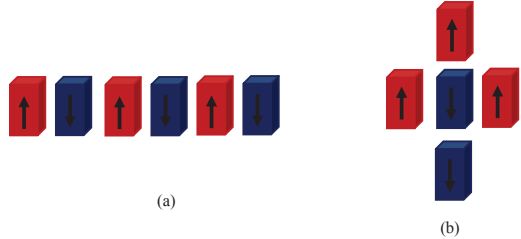}% Here is how to import EPS art
\vspace{-0.2in}
\caption{\label{fig2} (a) Inverter  (b) Majority gate.}
\end{figure}

Even though MCA can operate at room temperature, bit stability in the presence of thermal fluctuations put limitation on the size of the nanomagnet~\cite{mca_temp}. A minimum size requirement to prevent thermal fluctuations must be more than 20 nm.
 MCA logic device consists of a finite number of magnetic cells arranged in a specific fashion to accomplish computation. FIG. 2 illustrates the two basic building blocks that are used to construct MCA circuits. In this paper a wire or interconnect is a line of nanomagnetic cells that are coupled in an anti-ferromagnetic fashion with their neighboring elements. A MCA anti-ferromagnetic wire can provide an inverted result as shown in  FIG.~\ref{fig2}(a) which is determined by the number of nanomagnets in the array. The basic logic gate in MCA is majority gate as shown in FIG.~\ref{fig2}(b). The output of majority gate is dependent on the majority of the 3-input. The same majority gate can be used to perform AND or OR operation, by setting one input of the gate to logic ’0’ or ’1’, respectively. These fundamental building blocks have been experimentally demonstrated ~\cite{cowburn2002, majority, javier, jap} and provide all the functionality necessary to construct any Boolean function.
\vspace{-0.2in}
\subsection{\label{sec:level3}Defect Study}
To accurately capture the reliability effects, defects have on a MCA system it is necessary to have a working clock. Regardless of the implementation, a circuit or system made from Cellular Automata devices will realistically require some kind of clock structure that directs the computation and provides gain. Since magnetic interactions are direction-insensitive, additional control (clock) is needed apart from an input to drive the information flow from input to output. The clock helps the system to overcome the energy barriers between metastable states and the ground state. 
E-QCA defect characterization has been extensively studied~\cite{tahoori,lombardi}, however very little work has been reported on MCA defect characterization. The main focus of this paper is on the defect characterization of magnetic cellular automata wire under a working clock. There are two preliminary defect studies under conventional adiabatic ordering scheme~\cite{niemier_defect, isqed}. Both studies are not comprehensive, since defect masking towards the output (point of testing) could not be considered as perfect ordering.

This study assumes a functionally correct array with perfect ordering and the flow of information~\cite{nmdc}, while in~\cite{niemier_defect, isqed} the defect free order was itself in question. It is important to note that, the actual implementation can vary~\cite{carlton} but as long as ordering is correct, the defect characterization conducted in this work would remain valid. Since the functional ordering was achieved by spatially varying field implemented by script on OOMMF package, this study includes a large category of single cell defect. Moreover all our defect data and ground truth (probability of defect) are found from in-house fabrication experience as opposed to the common defect assumption.
\section{\label{sec:fab}Types of Fabrication Defects}
\begin{figure}
\includegraphics[scale=0.43]{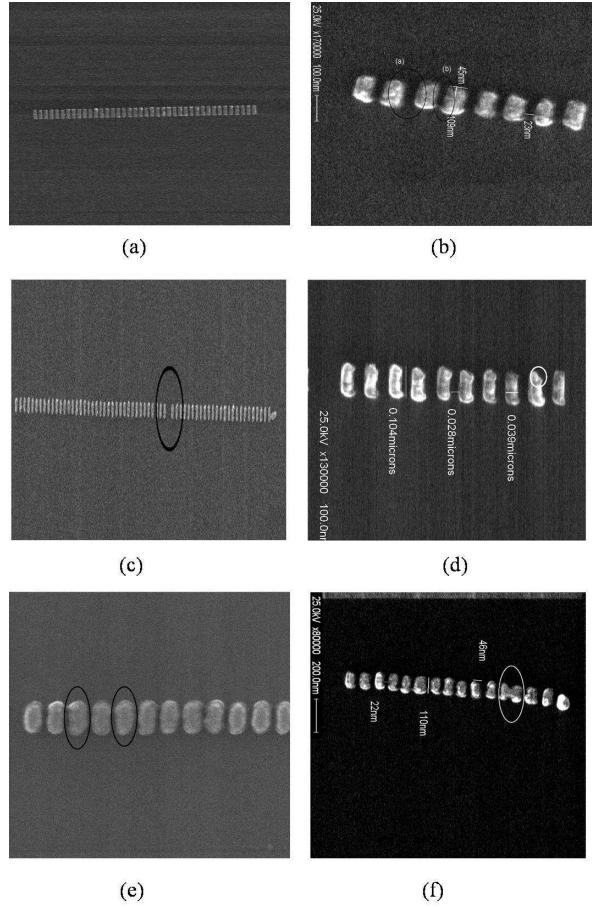}% Here is how to import EPS art
\caption{\label{fig3} An ideal chain, where the shape of the cells is regular. As a result the single domain dipole moments are strongly coupled.
(b) Irregular spacing between nanomagnets (A) Large spacing (B) Small
spacing. This can be due to blanking or deflection errors which may occur when the electron
beam is not deflected properly when it is supposed to or due to shaping errors which
 occur in variable-shaped beam systems. (c) Missing cell in an array, which may be due to unexposed of the resist caused by contamination of the resist. (d) Missing material defect (electrons from exposure of an adjacent region spill over into the exposure of the currently written feature, effectively enlarging its image result in merging). (e) Bulge defect seen in nanomagnet array and (f) Partial merge magnet in MCA array (electrons from exposure of an adjacent region spill over into the exposure of the currently written feature, effectively enlarging its image result in merging).}
\end{figure}
 In our Nano Research Center~\cite{NNRC}, we have successfully fabricated MCA arrays 
 of various lengths. The process begins with the coating of resist (PMMA) on Si
wafer, which is followed by lithography. We have used electron beam nano-lithography (JEOL SEM 840) with a high yield and NPGS system~\cite{npgs} to write the patterns on the resist, afterwards we deposited the ferromagnetic material of our choice, in this case permalloy, via the Varian Model 980-2462 Electron Beam Evaporator. We achieve a vacuum of about $2 \mu$ Torr and evaporated the material, after that we processed the lift-off to fabricate nano size magnets, with a considerable height and a good surface quality.
 FIG.~\ref{fig3}(a)
shows a defect free array, act as wire of 20 nm thick permalloy nanomagnets. Our main focus in this paper is to
characterize the defects in MCA array, which we have seen in our
experiments as shown in
FIG.~\ref{fig3}(a), (b), (c), (d), (e), (f).
When feature sizes in a device are small enough, the fabrication
defects in nano-fabrication methods can become a dominant factor which
determine the actual shape and operation of the nano-structure. The
cause of these defects can be extrinsic or intrinsic~\cite{2008nano}.
\begin{figure}[h]
\includegraphics[scale=0.25]{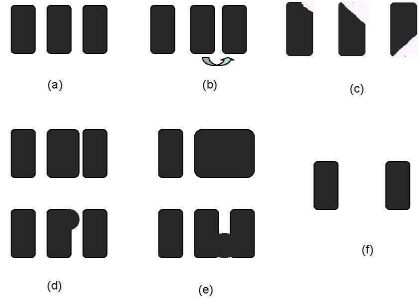}% Here is how to import EPS art
\caption{\label{fig4}Type of defect characterized (a) Defect free nanomagnets (base case) all the nanomagnets are of regular shape (50 x 100 x 20 nm$^3$)
with 20 nm uniform spacing between two nanomagnets, (b) Irregular spacing, two nanomagnet (2 and 3) comes close and other nanomagnet (magnet 1) moved farther apart, (c) Missing material defects with different amount of missing material and from different location (top/bottom) in nanomagnet, (d) Bulge defect with two form namely, uniform bulge (upper array) which results in an increase in the width of the $2^{nd}$ nanomagnet and non uniform bulge (bottom array) which results in the bump at the top of the $2^{nd}$ nanomagnet,
(e) Merge defect with two types, first fully merged cell as shown in upper array, magnet 2 and 3 merged completely and second is the partially merged cell (lower array) and 
(f) Missing cell defect, in which whole nanomagnet ($2^{nd}$) is missed.}
\end{figure}
We have noticed during our fabrication that defects may be possible in 
 the writing phase, in which the pattern is
transferred on the resist, and the deposition phase, in which
nanomagnets are deposited on the substrate. The cause of defects can be sample charging (either negative or
positive), backscattering calculation errors, dose errors, fogging
(long-range reflection of backscattered electrons), out-gassing,
contamination, beam drift and particles~\cite{nanolitho}. The essence of the observed defects in FIG.~\ref{fig3} are categorized into the following groups:
\begin{itemize}
\item[a)]{\textbf {Defect Free}}: all the nanomagnets are of regular shape of dimensions 50 x 100 x 20 nm$^3$ and optimum space of 
20 nm between nanomagnets, as shown in FIG.~\ref{fig4}(a).

\item[b)]{\textbf {Irregular spacing}}: the spacing between nanomagnets is not uniform (observed regularly). It is clear from the FIG.~\ref{fig4}(b) that two regular shaped nanomagnets (2 and 3) come closer which causes increase in the spacing between nanomagnets 1 and 2. In our experiments we have noticed that space irregularity is very common in MCA array.

\item[c)]{\textbf{Missing material}}: some portion of the particular nanomagnet is missing as compared to
 the original (defect free) arrangement as shown in FIG.~\ref{fig4}(c). In order to study
 missing material defect, 
we have considered two cases (1) Different amount of missing material as shown in FIG.~\ref{fig4}(c) where $1^{st}$ nanomagnet has less missing material as compared to the $2^{nd}$ and $3^{rd}$ nanomagnet (2) Different locations  (top/bottom) 
of missing material in nanomagnet as shown in FIG.~\ref{fig4}(c).
 
\item[d)]{\textbf{Bulge magnet}}: (a) uniform bulging in nanomagnet results in a big nanomagnet which in 
turn decreases the spacing  between two nanomagnets and hence also leads to irregular spacing as shown in FIG.~\ref{fig4}(d), top array ($2^{nd}$ magnet is bulged) (b) non-uniform bulge, in which there is an increase in the magnetic material which looks like bump as shown in FIG.~\ref{fig4}(d), bottom array. Note that all bulge defects are frequently observed.
\item[e)]{\textbf{Merged Magnet}}: Individual nanomagnet merge with the neighboring nanomagnet as shown in FIG.~\ref{fig4}(e). Further classifications are (a) fully merged cell, where two nanomagnets merge completely and form one big nanomagnet as shown in FIG.~\ref{fig4}(e), top array (b) partially merged cell, in which portion of the near-by two nanomagnets merge as shown in FIG.~\ref{fig4}(e), bottom array.
\item[f)]{\textbf{Missing cell}}: entire cell is missing (rarely occurring) as shown in FIG.~\ref{fig4}(f).
 \end{itemize}
In this framework, we have studied the geometry defect, which occurred during fabrication process. The geometry defects are deterministic in nature unlike the soft temporal error arising out of noise, stray magnetic field and temperature variations. The scope of the current study is (1) single occurrence of each defect type in an array and (2) assumption that  input-output nanomagnets are defect free. 

In our study, we have assumed room temperature operation (T=300K) and neglecting thermal fluctuations. Furthermore, the impact of stray noise and thermal fluctuation would be minimal at the dimensions of our study (as the ground state energy of the magnets under study (100 x 50 x 20 $nm^3$) are a few orders of magnitude of kT at room temperature ~\cite{csaba_power}). 

Also, in our study surface roughness is neglected and we have assumed the uniform crystalline structure.In this framework OOMMF tool by NIST has been used. The OOMMF has great convergence with quantitative SEMPA measurements.
\section{Defect Analysis}
We analyze the magnetic cells using Landau-Lifshitz equation (micro-magnetic theory) which accounts for various energies like Zeeman energy,
Magneto-static energy, Exchange energy, Anisotropy energy, Demagnetization energy to model the behavior of the defective array.
This framework presents an extensive and thorough defect study of realistic fabrication defects.  Because of the nano scale size magnet and inherent noise of electron beam lithography, the type of defect observed in fabrication process are 1) less frequent missing cell 2) minimum fully merge cell and 3) missing material, bulging and irregular spacing defects which occur commonly.
\begin{figure}
\includegraphics[scale=0.29]{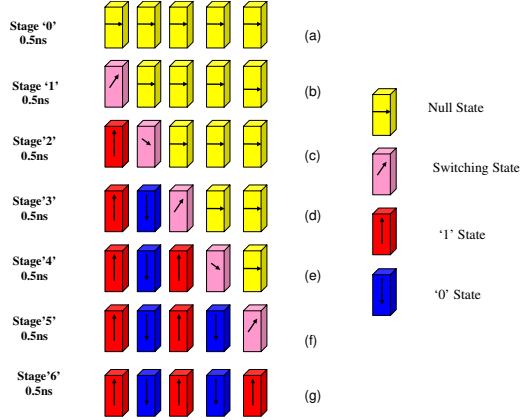}% Here is how to import EPS art
\caption{\label{landau_clock} Showing 7 stages for 5 nanomagnet chain. Yellow color shows the null state, Pink
color is for switching state, Red for logic one and logic zero for Blue. Each stage is for 0.5 ns: (a) Null state, all the nanomagnets in the array forced in hard (x-axis) axis. (b) The Null-field from the left most nanomagnet is moved, comes under the switching field, at this time input (logic one) is given. (c) 
Null-field from the second nanomagnet is moved, and comes under the influence of switching field while
remaining nanomagnets are at null-field. (d) The second nanomagnet attained the stable state according 
to the previous neighbor ($1^{st}$ nanomagnet) and third nanomagnet comes under the switching-field. }
\end{figure}
\subsubsection{Micro-magnetic Simulation Parameters}
We performed the micro-magnetic simulation using the OOMMF code, based on the Landau-Lifshitz equation.
The program approximates the continuum micro-magnetic theory, where the magnetic sample was divided into a regular two dimensional grid of square cells. Within each cell, the magnetization is assumed to be uniform and is represented by a three dimensional spin vector M. The exchange energy is computed via eight-neighbor scheme~\cite{donahue}, while the magneto-static energy is calculated via a fast Fourier convolution of the magnetization.
The dimension for nanomagnets used for this analysis is 50 x 100 x 20 $nm^3$,
and spacing between nanomagnets is 20 nm. A unit cell size of 5 nm was used for the simulations.
The parameter values used for the numerical calculations were characteristic of permalloy ($Ms$ = $85 \times 10^5$~$ A/m$,~$A $~= $1.3 \times 10^{-12}$~ $J/m$,~ $K$~ = $500$~$ J/m^3$).

 We commence this section by explaining the simulation process, wherein we have considered an array of 8 nanomagnets and
16 nanomagnets of MCA. 
In this frame of work, we have used a spatially moving clock scheme~\cite{nmdc} for defect analysis in MCA
architecture, where we deliver a spatially moving clock from input to output. The information processing  requires the interaction of the three fields: null field, switching field and input field.  In this, the null field (clock) of 100~$mT$ along the hard-axis (in x direction) was applied which forced all
 the magnetic cell moments to align in the direction of field (hard-axis) and magnetization in easy-axis reached zero.
As we vary field spatially, the nanomagnets come under the influence of switching field (30~$mT$) and finally start aligning anti-ferromagnetically, according to the input magnetization (Fig.~\ref{landau_clock} illustrates the sequence of the events). 
\begin{figure}
\includegraphics[scale=0.25]{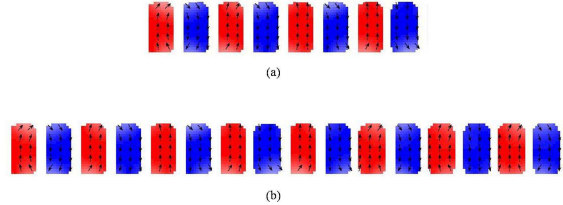}% Here is how to import EPS art
\caption{\label{fig5} Defect free nanomagnet array: (a) 8 nanomagnet MCA array (b) 16 nanomagnet MCA array}
\end{figure}
 The propagation happens primarily due to the
neighboring cell exchange coupling and shape-anisotropy. The simulation result
in FIG.~\ref{fig5}, under no defect, shows that magnetic arrays behave
perfectly, reaching near-uniform magnetization of individual cells, arranged in perfect anti-ferromagnetic order.
\begin{figure}
\includegraphics[scale=0.25]{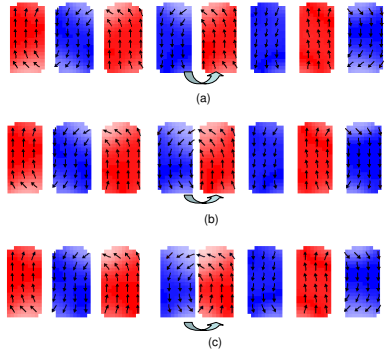}% Here is how to import EPS art
\caption{\label{fig6} Irregular spacing defect in 8 nanomagnets: (a) Irregular space (5 nm) between nanomagnets 4 and 5, (b) Irregular spacing of 10 nm and (c) Irregular space between nanomagnets 4 and 5 of 15 nm, all defects are masked.}
\end{figure}
\subsubsection{Single Irregular Spacing Defect}
This is the most noticeable fabrication defect. Here we want to clarify that this
study is not meant to decide optimum spacing. We have concluded that
$20$ nm spacing is optimum for our magnet dimensions which also confers
with the relative dimensions used by Csaba~{\it
et. al.},~\cite{Csaba003}. In this work, we intend to study effect of a single regular shaped cell spaced irregularly
due to fabrication variations (caused by effects like stray magnetic
fields, thermal fluctuations, contaminated resist etc.) common in
litho-based assembly. We need a robust architecture that would be
tolerant to small spacing irregularities. 
 FIG.~\ref{fig6} shows, nanomagnets 4 and 5 spaced irregularly.
 Irregular space defect has been studied with conventional clock~\cite{niemier_defect}, wherein the irregularity in space cause stuck-at-fault.
 However our analysis suggested that irregular spacing does not affect the functionality of the MCA array. In FIG.~\ref{fig6}, 
 MCA array of 8 nanomagnet is simulated with different space irregularities (25\%, 50\%, 75\%) and 
observed that MCA arrays are more tolerant against irregular space defects. 
\subsubsection{Missing Material Defect}
We have characterized the missing material defect in two different forms, namely (1) Different amount of missing material and (2) Different location (top/bottom) of missing material in MCA arrays. 
\begin{figure}
\includegraphics[scale=0.2]{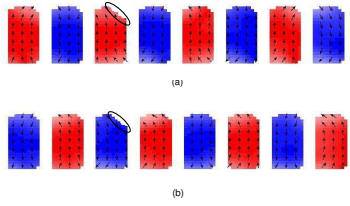}% Here is how to import EPS art
\caption{\label{fig7}Missing material less than 5\% in $3^{rd}$ nanomagnet  in 8  nanomagnet array, gives correct output.}
\end{figure}

\begin{figure}
\includegraphics[scale=0.28]{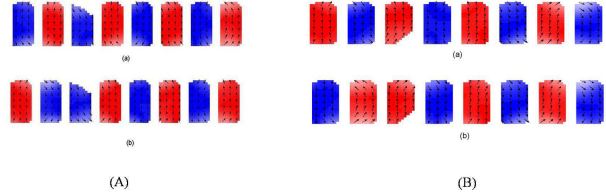}% Here is how to import EPS art
\caption{\label{fig8}(A)~Missing material with large amount from top of the $3^{rd}$ nanomagnet (a) logic zero propagate correctly (b) logic one does not propagate correctly as the $3^{rd}$ nanomagnet did not flip and retains old value. (B) Missing material from bottom of the $3^{rd}$ nanomagnet.}
\end{figure}
First set of simulations dealt with the small amount of missing material defect. The 8 nanomagnet MCA array was considered and  5\% of material from top of the nanomagnet was removed as shown in FIG.~\ref{fig7}. 
The array displayed perfect logics (`1' and `0') with small material loss.
Simulations were repeated again but now with the removal of high percentage of missing  material (20\%).
Subsequently, we found that if the material is missing from the top, 
the magnetization is always downwards in the defective nanomagnet as shown in FIG.~\ref{fig8}~(A), 
whereas if the material is removed from the bottom, the magnetization is always upward (logic 1) as shown in FIG.~\ref{fig8}~(B).
Shape anisotropy property of single domain nanomagnet plays an important role (as the perfect part has better defined shape as compared
 to the missing material part). As the percentage of missing material 
increased, it was difficult to null the nanomagnet, hence after removal of clock field the nanomagnet would retain the old value.
Thus, the missing material more than 5\% results in the stuck-at-0 or stuck-at-1 depending on the location of the missing material (top or bottom).
\subsubsection{Single Bulge Defect}
\begin{figure}
\includegraphics[scale=0.45]{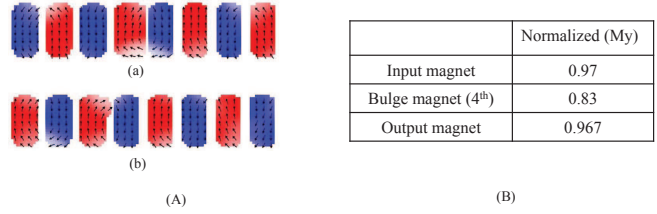}% Here is how to import EPS art
\caption{\label{fig9}(A) Results for bulge defect in 8 nanomagnet array (a) uniform bulge at $4^{th}$ nanomagnet  (30\% increase in the width of the nanomagnet) results in the decrease in the spacing between $4^{th}$ and $5^{th}$ nanomagnet (b) non-uniform bulge at the $3^{rd}$ nanomagnet (small bump at the top) (B) Normalized magnetization in nanomagnet.}
\end{figure}

We have seen in our fabrication that the probability of occurrence 
of bulge in nanomagnet is very high due to unavoidable lithography variations. We have characterized the bulging defect in 
two scenarios as discussed in previous section.
The study begins with the uniform bulging in nanomagnet resulting in increase in the width of the nanomagnet
which causes the decrease in the spacing between the neighboring nanomagnet (left or right).  Niemier {\it{et.al.}} have demonstrated the effect of bulged nanomagnet array with conventional clock and reported that bulging results in weaker �1� and �0�. In this frame of work we have explored the bulge defect. We have assumed only one bulge defect in an array as shown in FIG.~\ref{fig9}. 
We have simulated the 8 nanomagnet array with different percentages of bulge defects (10\%, 20\%, 30\%). FIG.~\ref{fig9}~(A)(a) 
shows the results for 30\% bulge (causing increase in width from 50 nm to 65 nm and decrease in spacing from 20 nm to 5 nm) in the nanomagnet. 
This is the maximum bulge in the nanomagnet because more bulge will result in the merging with the neighboring cell and cause merge cell defect which we will be discussing in following section.
We have observed that magnetization in bulge nanomagnet ($4^{th}$) is weak in Y-axis as shown in FIG.~\ref{fig9}~(B). However, the output magnetization is restored. Hence, due to shape anisotropy if the output nanomagnet is defect free, the bulge in the nanomagnet would not have any effect on the correct flow of information.
 Next we simulated the array with non-uniform bulging, commonly encountered in our fabrication experiments.  FIG.~\ref{fig9}~(A)(b) shows the results
for non-uniform bulge at 3$^rd$ nanomagnet (top) in 8 an MCA array. The non-uniform bulge (at top of the $3^{rd}$ nanomagnet) causes decrease in the space between
nanomagnet 3 and 4 at the top. 
 It is evident from the results that non-uniform bulging does not affect the correct flow of information as shown in FIG.~\ref{fig9}~(B)~(b).
\begin{figure}
\includegraphics[scale=0.2]{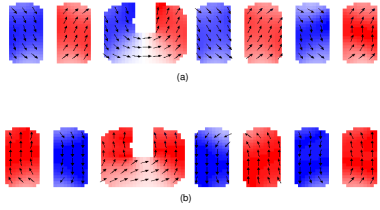}% Here is how to import EPS art
\caption{\label{fig10}Partially merged cell defect in 8 nanomagnet array, nanomagnet 3 and 4 merged, (a) Initial state (b) Logic one cause error because  the nanomagnet after merged cell does not flip and retains the old value.}
\end{figure}

\begin{figure}
\includegraphics[scale=0.2]{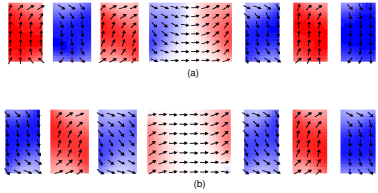}% Here is how to import EPS art
\caption{\label{fig11} Merge cell defect in 8 nanomagnet array, nanomagnet 4 and 5 merged, (a) logic one propagates correctly but (b) logic zero cause error because the merged nanomagnet does not flip.}
\end{figure}
\subsubsection{Merged Neighboring Cell Defect}
We have seen two types of merge cell defect.  First, the partial merge cell defect, in which portion of the near-by two nanomagnets are merged as shown in FIG.~\ref{fig10}.  Second, the fully merged cell, where two nearby nanomagnets are merged completely (forming one big nanomagnet) as shown in FIG.~\ref{fig11}. We have simulated the effect of merge cell defect with spatially moving clock. 

First set of simulations was run for partial merge cell defect. It is illustrated in  FIG.~\ref{fig10} that nanomagnet 3 and 4 partially merged, where FIG.~\ref{fig10}~(a)  is the initial state (logic '0' propagation). 
We have applied the spatially moving clock in hard axis and input (logic '1') to the left most nanomagnet.
The results shows that nanomagnet after the partial merge defect does not flip 
(retains the old value) which cause incorrect output as shown in FIG.~\ref{fig10}~(b).

The second set of simulation involves the Merge cell defect as shown in FIG.~\ref{fig11} where nanomagnet 4 and 5 merged completely, which results in one big nanomagnet. FIG.~\ref{fig11}~(a) is the initial state (logic one propagation). We applied the clock in hard axis and input (logic zero) was  applied to left most nanomagnet. The results shows that Merge cell defect causes incorrect output as shown in FIG.~\ref{fig11}~(b).
 This is due to the merge cell having more magnetization in the x-axis as compared to the y-axis, therefore it does not flip. The nanomagnet after merge cell retains the previous value. Hence array works
well with one input (logic `0' or logic `1') depending on the initial state of the array.
\subsubsection{\label{sec:missing cell}Missing Cell Defect}
\begin{figure}
\includegraphics[scale=0.2]{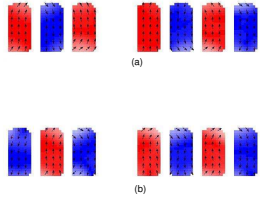}% Here is how to import EPS art
\caption{\label{fig12}Missing  nanomagnet in 8 nanomagnet array. (a) Logic one propagate, (b) Logic zero cause incorrect output.}
\end{figure}
A large number of missing cell study is performed in the context of E-QCA~\cite{tahoori, lombardi, bhanja_qca}.
 In our fabrication experiments, 
we have seen some missing cell as shown in FIG.~\ref{fig3}(c) in rare occasions.
 The MCA array behaves as decoupled arrays due
 to the large spacing created by the missing cell. MCA array with missing cell (4th nanomagnet) is shown in FIG.~\ref{fig12}.
The initial state of MCA array with missing cell is illustrated by FIG.~\ref{fig12}(a). We have applied input to the left
 most nanomagnet and spatially moving clock in hard-axis of the nanomagnets. 
The input propagated correctly till 3rd nanomagnet and since 4th nanomagnet is missing, the 5th nanomagnet has 
no influence from the 3rd nanomagnet due to large spacing. Hence, the fifth nanomagnet will retain its old value as shown in FIG.~\ref{fig12}(b). So it will propagate
 either logic '1' or '0' depending on the initial state of the nanomagnet after the missing cell. 
Therefore, there is a 50\% probability of getting correct output.
													
\subsubsection{Role of array length in defect}
\begin{figure}
\includegraphics[scale=0.2]{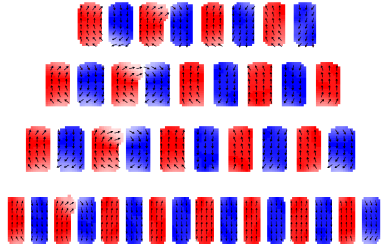}% Here is how to import EPS art
\caption{\label{fig13} Bulge defect at nanomagnet 3 in 8, 9, 10, 16 nanomagnet MCA array.}
\end{figure}

We have studied these defects with different array lengths.
FIG.~\ref{fig13} shows the non uniform bulge defect at $3^{rd}$ nanomagnet in 8, 9, 10 and 16 nanomagnet MCA array. 
It is evident from the results that it is working perfectly irrespective of the length. 
This is a feature of spatially moving clock, where nanomagnet in switching state is influenced by the stable state of the previous nanomagnet as opposed to the following
nanomagnet, which is in null state. This is because the magnetization of the null state nanomagnet is in X-axis. The null field is removed from all the nanomagnets one by one, resulting in directional information flow from input to output. Hence length is not a determinant in defect behavior.

\subsubsection{Role of location of defects in array}
\begin{figure}
\includegraphics[scale=0.2]{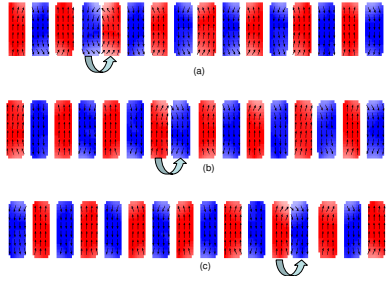}% Here is how to import EPS art
\caption{\label{fig14} Irregular spacing defect in 16 nanomagnets: (a) Irregular space (5 nm) between nanomagnet 4 and 5, (b) Irregular spacing between nanomagnet 8 and 9,  (c) Irregular space between nanomagnet 12 and 13. All the defects are masked.}
\end{figure}

Here, we focus our attention on the effect of the defect location in MCA array. 
We have assumed that the input and output nanomagnets are defect free and there is a single irregular spacing defect in an array. Through our simulation experiments we observed that the location of defect in an array does not affect the correct propagation flow of information.
 The simulation results for irregular spacing defect in 16 MCA array at three different locations is shown in FIG.~\ref{fig14}.
\begin{table*}
\begin{center}
{\footnotesize 
\begin{tabular}{c }
 \\
%\begin{tabular}{p{2.5in} p{0.9in}p{1.0in}}  \hline
\begin{tabular}{p{2.0in}p{1.6in}p{1.2in}}  \hline
&\multicolumn{2}{c} {$R_{D_{i}}$} \\
Defect type & 8   nano-magnets & 16 nano-magnets \\ \hline   
Bulge defect & 100\%& 100\% \\ 
Missing material (less than 5\%) & 100\%& 100\% \\ 
Missing cell & 50\%& 50\% \\ 
Merged cell & 50\%&  50\% \\
Irregular spacing & 100\% & 100\% \\ \hline
\end{tabular} 
\end{tabular}
}
\end{center}
\caption{Robustness measures for 8 and 16 MCA arrays with respect to various types of defects.  }
\label{rob}\end{table*}
\subsection{Discussion}
The important points observed by our simulation experiments are summarized below:
 \begin{enumerate}
\item MCA systems are robust towards most of the defects (irregular spacing, bulge, missing material (5\%)).
\item The location of a defect in an array does not affect the defect masking.
\item The stuck-at fault in missing material defect depends on the position of the missing material (top/bottom) in nanomagnet in MCA array, because in top missing material the magnetization in Y-axis is always downward while in bottom missing material the magnetization is always upwards.
\item The stuck-at fault in a missing cell depends on the initial polarization of the MCA array. The nanomagnet after the missing cell does not have any effect from previous nanomagnet (due to large space created by missed cell), hence retains the old (initial) value.
\item Also, stuck-at-faults in a merge cell depends on the initial polarization of the MCA array. In merged nanomagnet the magnetization in x-axis is more than y-axis so it is hard to flip the merged nanomagnet. Hence nanomagnet retains the old (initial) value which  propagate either logic'0' or logic'1'.
\end{enumerate}
\subsection{Output Magnetization under Defect}
\begin{figure*}
\includegraphics[scale=0.4]{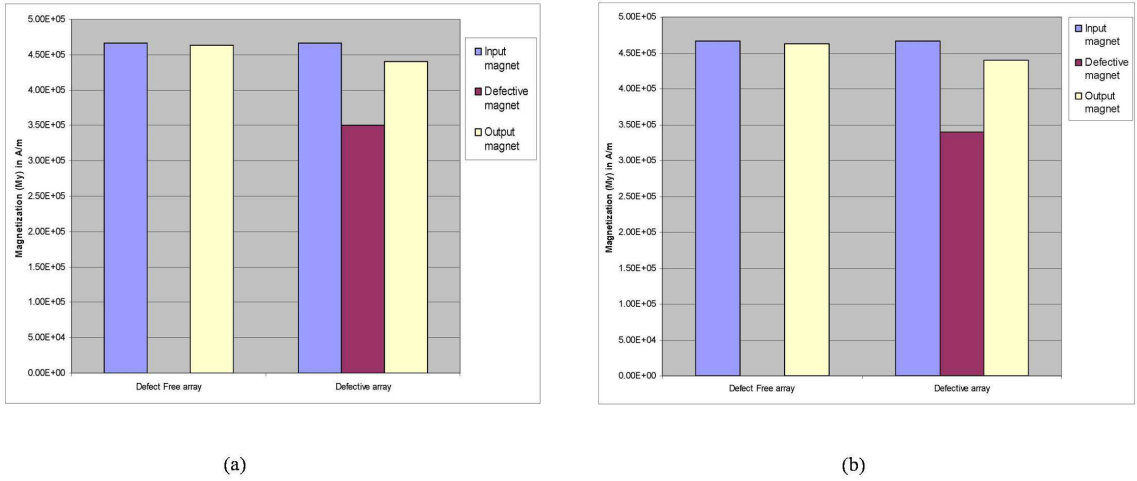}% Here is how to import EPS art
\caption{\label{fig15}  (a) Measured magnetization for defect free and defective 8 MCA array (b) Measured magnetization for defect free and defective 16 MCA array.}
\end{figure*}
We have measured the input and output magnetization for the MCA array 
of length 8 and 16 nanomagnets as shown in FIG.~\ref{fig15}. It is evident from the FIG.~\ref{fig15} that output magnetization strength is uniform for 
both the arrays. Next, for defective array (missing material defect), we observed 
that the magnetization of the defective nanomagnet is weak as shown in FIG.~\ref{fig15} (maroon bar). However
 the defect in the array does not affect significantly the strength of the output magnet irrespective of the type of defect. 
This is no surprise to us as shape anisotropy in single domain nanomagnet is inherent to MCA architecture, which helps restore the magnetization of subsequent magnet.
At this point, we introduce an index of defect-tolerance, $R_{D}$. This is the
 robustness with respect to the type of defect.

\begin{eqnarray}
R_{D_{i}} & = & p(correct~output|Defect_{i}~ has~ occurred) \nonumber 
\label{robust}
\end{eqnarray}
It is clear from Table.~\ref{rob} that Missing material defect (less than 5\%), bulge nanomagnet defect and
 irregular space defect for array length 8 and 16, irrespective of the location of defect is extremely robust. Merge cell defect and Missing cell defect, which have less probability of occurence have 50\% chance of correct output. Hence the MCA architecture is more robust.
\section{conclusion}
 We have studied the defect in MCA array for robustness of nano-magnetic arrays based on the theoretical framework of Landau-Lifshitz, capturing many effects like, shape anisotropy, crystalline anisotropy, exchange coupling between
neighboring cells. First, we observed that with spatially varying clock scheme, length of an array and location of defect does not affect the output. Secondly, there was no signal degradation over the length of array and finally, we demonstrated that with spatially moving clock, MCA  architecture is extremely robust to irregular space, bulge defect and missing material(less than 5\%), which are ordinarily encountered in our fabrication experiments. Our future study include probabilistic defect macro-modeling.

%\nocite{*}
%\bibliographystyle{}
\bibliography{M-QCA,M-QCA1}

\end{document}